\documentclass[11pt]{article}
 \usepackage{graphicx,citesort}
  \usepackage{amssymb,amsmath,bm}
\makeatletter
\renewcommand{\section}{\@startsection{section}{1}{0pt}%
{-3.5ex plus -1ex minus -.2ex}{2.3ex plus.2ex}%
{\normalfont\large\bfseries}}
\def\postsection{.\@postskip@}
\makeatother

\begin{document}

\begin{center}
{\LARGE\textbf{Resummation in (F)APT}}
 \vskip7.5mm

{\large A.~P.~Bakulev and S.~V.~Mikhailov}
 \vskip5mm

\textit{Bogoliubov Lab. Theor. Phys., JINR\\ 
        Jolio-Curie 6, 141980 Dubna, Russia}
 \vskip10mm

\textbf{Abstract}
 \vskip2.5mm

\parbox[t]{110mm}{{\small
 We present new results on the summation of nonpower-series expansions 
 in QCD Analytic Perturbation Theory (APT) in the one-loop approximation.
 We show how to generalize the approach 
 suggested by one of us earlier to the cases 
 of APT and Fractional APT (FAPT) 
 with heavy-quark thresholds.
 Using this approach we analyze 
 the Higgs boson decay $H^0\to\bar{b}b$.
 We produce estimations of the higher-order corrections importance
 in (F)APT 
 and present a very transparent interpretation 
 of the resummation results
 in terms of $\Lambda_\text{QCD}$ shifts.
}} 
\end{center}
 \vskip5mm

\section{Introduction}
\label{sec:intro}

This is a written version of the talk
given at the Memorial Igor Solovtsov Seminar,
held on January 17--18, 2008, 
at the Bogoliubov Laboratory of Theoretical Physics (BLTPh),
JINR (Dubna).
For this reason first we want to recollect 
our contacts with Igor Solovtsov in and outside the lab.

We met first time at BLTPh in the mid 90s 
when he conducted a series of seminars on the variational approach
to quantum field theory, 
in particular to QCD.
He was those times very enthusiastic about this approach
which later on guided him to Analytic Perturbation Theory.
Only after some time other people started 
to share his enthusiasm and understand 
what he knew at the very beginning.
We met ones at the Institute for Theoretical Physics of 
Heidelberg University (presumably, in 1997) 
and participated 
in the seminar on Analytic Perturbation Theory 
he presented there --- it was a kind of enjoinment,
so clearly he presented the matter. 
We remember very well how he was preparing his doctorate thesis
--- calmly, with some kind of dignity, 
and remaining to be very open for discussions.
Later on, we met also during some conferences. 
The one we remember more clearly, 
is the Research Workshop on Calculations for Modern and Future Colliders,
held in Dubna in July 2000.
The reason for this memory is especial:
during the boat-trip along the Volga river
it appeared that Igor was a very good singer --- 
he had a good voice, an absolute pitch 
and a very nice song repertory.
One of us, A.B., 
met Igor last time quite recently, 
during the XIV International Seminar 
``Nonlinear Phenomena in Complex Systems'' 
(May 22--25, 2007, Minsk) and ones again enjoyed his nice voice.
To resume, Igor was a talented physicist, a nice man, and a good friend.
Let God bless his memory and Igor be remembered forever.

Now we continue with his `child'
--- APT.
Since the original works of Jones and Solovtsov~\cite{JS95-349,JS95-357} 
and Shirkov and Solovtsov \cite{SS96} 
appeared in 1995--97, 
the analytic approach to QCD perturbation theory 
has progressed considerably.
The main object of this approach is the spectral density 
that provides the means to define 
the analytic running coupling 
in the Euclidean region using a dispersion relation.
The same spectral density is used to define also 
the running coupling in Minkowski space, 
by employing the dispersion relation for the Adler function 
\cite{MS96,MiSol97}.
The analytic approach was extended beyond 
the one-loop level \cite{MiSol97,SS98} 
and special analytic and numerical tools were elaborated 
\cite{Mag99,KM01,Mag05}.
APT appeared to be well-suited  to deal with more than one 
(large) momentum scale~\cite{SSK99,BPSS04}.
All these efforts resulted in a systematic approach, 
named Analytic Perturbation Theory (APT), 
recently reviewed in \cite{SS06}.

In order to treat hadronic amplitudes in the same manner
one needs to use not only 
analytization of the running coupling,
but also more complicated objects,
like integrated couplings
$\int_{0}^{1}\!dx\!\int_{0}^{1}\!dy\,
  \alpha_{s}\left(Q^{2}xy\right) f(x)f(y)$,
  as was firstly indicated in~\cite{KS01,Ste02}.
The authors of this paper proposed the principle of analytization 
\textit{as a whole} 
to treat such objects 
and new techniques was developed 
to produce needed analytic images. 
The same principle applies also 
to other hadronic amplitudes in the renormalization-group based 
QCD factorization approach,
where complicated objects,
like fractional powers of running coupling 
and its products with logarithms, $\ln(Q^2/\mu^2)$,
naturally arise.
A version of the ``analytization'' approach,
which takes into account all terms 
that contribute to the spectral density,
was developed in~\cite{BMS05,BKS05,BMS06}
and named 
Fractional Analytic Perturbation Theory (FAPT).

In this paper, 
we present new results on the resummation 
of non-power-series expansions in APT in the one-loop approximation.
We show how to generalize the approach 
suggested by one of us earlier~\cite{MS04} 
to the cases of: 
(i) global (with taking into account heavy-quark thresholds) APT, 
and
(ii) global FAPT.
Using this approach we analyze  
the Higgs boson decay into a $\bar{b}b$-pair of quarks.
We produce estimations of how important 
higher-order corrections can be 
and present very transparent interpretation 
of the resummation results in terms of $\Lambda_\text{QCD}$ shifts.

\section{Fixed-$N_f$ and global APT}
\label{sec:APT.Summation}
First, we need to explain our notation.
In order to have a direct connection to our previous papers~\cite{BMS05,BKS05,BMS06}
and to simplify the main formulae in the sections,
where we speak about fixed-order (F)APT with fixed value of active flavors $N_f$,
we use here the normalized coupling~\cite{MS04}
$a(Q^2)$=$\beta_f\,\alpha_\text{s}(Q^2;N_f)$
and analytic images are constructed for powers of this object: 
\begin{subequations}
 \label{eq:Glo.Couplings}
\begin{eqnarray}
 {\cal A}_{n}(Q^2) 
  &=& \textbf{A}_\text{E}\left[a^{n}_{}\right]
  \ \equiv\
   \int_0^{\infty}\!
    \frac{\rho_n(\sigma)}
         {\sigma+Q^2}\,
      d\sigma\,;
 \label{eq:A.rho.Nf}\\
 {\mathfrak A}_{n}(s)
  &=& \textbf{A}_\text{M}\left[a^{n}_{}\right]
  \ \equiv\
    \int_s^{\infty}\!
     \frac{\rho_n(\sigma)}
          {\sigma}\,
      d\sigma~~~
 \label{eq:U.rho.Nf}
\end{eqnarray}
with $\rho_n(\sigma)=\textbf{Im}\,\left[a^{n}(-\sigma)\right]$
and $\beta_f=b_0(N_f)/(4\pi)=(11-2N_f/3)/(4\pi)$ 
with $b_0(N_f)$ 
being the first coefficient in the QCD $\beta$ function.

When we discuss a global version of (F)APT,
where $Q^2$ (or $s$) varies in the whole domain $[0,\infty)$ 
and $N_f$ effectively becomes dependent on $Q^2$ (or $s$),
we use the standard running coupling $\alpha_\text{s}(Q^2)$, 
and analytic powers 
are defined for the corresponding powers of this coupling
\begin{eqnarray}
 {\cal A}^\text{\tiny glob}_{n}(Q^2) 
  = \textbf{A}_\text{E}\left[\alpha_{s}^{n}\right]
  \ \equiv\
    \int_0^{\infty}\!
     \frac{\rho^\text{\tiny glob}_n(\sigma)}
          {\sigma+Q^2}\,
       d\sigma\,;
 \label{eq:A.n.Glo}\\
 {\mathfrak A}^\text{\tiny glob}_{n}(s)
  = \textbf{A}_\text{M}\left[\alpha_{s}^{n}\right]
  \ \equiv\
    \int_s^{\infty}\!
     \frac{\rho^\text{\tiny glob}_n(\sigma)}
          {\sigma}\,
      d\sigma\,.~
 \label{eq:U.n.Glo}
\end{eqnarray}
In order to distinguish these two approaches,
we introduce the upper index $^\textbf{\tiny glob}$.
Here we need also analogs of fixed-$N_f$ quantities with
standard normalizations:
\begin{eqnarray}
 \bar{\cal A}_{n}(Q^2;N_f) 
  = \frac{{\cal A}_{n}(Q^2)}{\beta_f^n}\,;\quad
 \bar{\mathfrak A}_{n}(s;N_f) 
  = \frac{{\mathfrak A}_{n}(s)}{\beta_f^n}\,,~
 \label{eq:Bar.Couplings}
\end{eqnarray}
which correspond to analytic couplings 
${\cal A}_{n}$ and ${\mathfrak A}_{n}$
in the Shirkov--Solovtsov notation~\cite{SS06}.
These couplings also have integral representations 
of the type (\ref{eq:A.rho.Nf})--(\ref{eq:U.rho.Nf})
with the spectral densities
$\bar{\rho}_n(\sigma;N_f)={\rho}_n(\sigma)/{\beta_f^n}$.

The spectral density $\rho^\text{\tiny glob}_n(\sigma)$
is simply related with the spectral density
$\bar{\rho}_n(\sigma;N_f)$~\cite{SM94,SS96,Mag99}:
\begin{eqnarray}
 \rho^\text{\tiny glob}_n(\sigma) 
  = \theta\left[\sigma\leq m^2_{4}\right]\,
     \bar{\rho}_n(\sigma;3)
  + \sum\limits_{f=4}^{6}
     \theta\left[m^2_{f}<\sigma\leq m^2_{f+1}\right]\,
      \bar{\rho}_n(\sigma;N_f)\,,
 \label{eq:rho_n.Glo}
\end{eqnarray}
\end{subequations}
where,
in order to make our formulae more compact,
we denoted $m_4=m_c$, $m_5=m_b$, $m_6=m_t$, 
and $m_7=+\infty$.

We need also to speak about effective coupling as a function
not of $Q^2$ or $s$, 
but rather of logarithms $L=\ln(Q^2/\Lambda^2)$ or $L=\ln(s/\Lambda^2)$.
In these cases, we use the same notation for coupling
but with an argument placed in squares:
$a^\nu[L]$, ${\cal A}_\nu[L]$, and ${\mathfrak A}_\nu[L]$.
Then, in the one-loop approximation we have
\begin{eqnarray}
 \label{eq:a.1-loop}
  a[L]
   = \frac{1}{L}\,,\quad
  \rho_1[L]
   = \frac{1}{L^2+\pi^2}\,,\quad
  \bar{\rho}_1[L;N_f]
   = \frac{1}{\beta_f\left[L^2+\pi^2\right]}\,.  
\end{eqnarray}

\subsection{Series summation in the one-loop APT with fixed $N_f$}
\label{subsec:APT.Nf.Fixed}
Let us consider different sorts of perturbative series expansions 
of a typical phys\-ical quantity, 
like the Adler function, $D[L]$,  
and write 
at the one-loop level \cite{BMS06},
\begin{eqnarray}
 \label{eq:Ini.Series}
 \left\{
\begin{array}{l}
  D[L]\\
{\cal D}[L]\\
 {\cal R}[L]
 \end{array}
 \right\}  = d_0+d_1\sum_{n=1}^{\infty} \tilde{d}_n\,\left\{
\begin{array}{l}
  a^{n}[L]\\
{\cal A}_{n}[L]\\
 {\mathfrak A}_{n}[L]
 \end{array}
 \right\} \,,
\end{eqnarray}
where
$L=\ln\left(Q^2/\Lambda^2\right)$
applies to the Euclidean ($D[L]$, ${\cal D}[L]$, $a^n[L]$ and ${\cal A}_n[L]$)
and
$L=\ln\left(s/\Lambda^2\right)$
to the Minkowski (${\cal R}[L]$ and ${\mathfrak A}_n[L]$)
regions,
and $\tilde{d}_n\equiv d_n/d_1$.
It is useful to introduce a generating function $P(t)$
for the series expansion
whose specific form depends 
on a particular quantity in question:
\begin{equation}
 \tilde{d}_n
  =\int_{0}^\infty\!\!P(t)\,t^{n-1}dt
   ~~~\text{with}~~~
   \int_{0}^\infty\!\!P(t)\,d t = 1\,.
 \label{eq:generator}
\end{equation}
To shorten our formulae, we use a shorthand notation
\begin{equation}
 \langle\langle{f(t)}\rangle\rangle_{P}
  \equiv
   \int_{0}^\infty\!\!f(t)\,P(t)\,dt\,,
\label{eq:average}
\end{equation}
where the brackets $\langle\langle{\ldots}\rangle\rangle_{P}$ denote 
the average over $t$ with the weight $P(t)$.

The different couplings $a$, ${\cal A}_{1}$, and ${\mathfrak A}_{1}$ 
satisfy the same one-loop renormaliza\-tion-group equation 
that can be rewritten
as a one-loop recurrence relation \cite{Rad82,Shi98,BRS00}
\begin{equation}
\left\{
\begin{array}{l}
  a^{n+1}[L]\\
{\cal A}_{n+1}[L]\\
 {\mathfrak A}_{n+1}[L]
 \end{array}
\right\}
 =
   \frac{1}{\Gamma(n+1)}\left( -\frac{d}{d L}\right)^{n}
\left\{
\begin{array}{l}
  a^{1}[L]\\
{\cal A}_{1}[L]\\
 {\mathfrak A}_{1}[L]
 \end{array}
\right\}\, .
\label{eq:recurrence}
\end{equation}
Substituting Eqs.\ (\ref{eq:recurrence}), (\ref{eq:generator}) into
the perturbative series expansion, Eq.\ (\ref{eq:Ini.Series}),
one obtains
\begin{eqnarray}
\left\{
\begin{array}{l}
  D[L]\\
{\cal D}[L]\\
 {\cal R}[L]
 \end{array}
\right\}
   &=& d_0 
   + d_1\sum_{n=0}^{\infty}
         \frac{\langle\langle{(-t)^{n}}\rangle\rangle_{P}}
              {n!}\,
          \frac{d^n}{dL^n}
           \left\{
            \begin{array}{l}
             a[L]\\
             {\cal A}_{1}[L]\\
             {\mathfrak A}_{1}[L]
            \end{array}
           \right\}\nonumber\\
   &=& d_0 + d_1\left\{
            \begin{array}{l}
             \langle\langle{a[L-t]}\rangle\rangle_{P}\\
             \langle\langle{{\cal A}_{1}[L-t]}\rangle\rangle_{P}\\
             \langle\langle{{\mathfrak A}_{1}[L-t]}\rangle\rangle_{P}
            \end{array}
           \right\}\,,~~~
\label{eq:integ-pres}
\end{eqnarray}
as it was shown by one of us (S.M.) in~\cite{MS04}.
As long as we have not proved that the order of summation and
integration can be interchanged, this representation has only
a formal meaning.
Note, however, that the integration over the Taylor expansion
of the term $a[L-t]$ in the integrand reproduces the initial
series for any partial sum.
It is not surprising that the integrand in the standard case
(first line in Eq.\ (\ref{eq:integ-pres})) 
faces a pole singularity
and is, therefore, ill-defined.
On the other hand, for the last two cases the integral has a
rigorous meaning, ensured by the finiteness of the couplings, i.e.,
${\cal A}_{1}[t]\; , {\mathfrak A}_{1}[t] \leq 1$.
Since any coefficient $\tilde{d}_n$ 
is the moment of $P(t)$,
this function should fall off faster than any power, 
e.g., like
an exponential or faster.
Therefore, the APT expressions on the RHS of Eq.\ (\ref{eq:integ-pres})
exist and are proportional to
${\cal A}_{1}[L-\bar{t}(L)]$ or ${\mathfrak A}_{1}[L-\bar{t}(L)]$,
where $\bar{t}(L)$ can be treated for each $L$ 
as an average value of $t$ 
associated with this quantity.

Provided the generating function $P(t)$ is known, 
one can compute the average $\langle\langle{{\cal A}_{1}[L-t]}\rangle\rangle_{P}$
(or $\langle\langle{{\mathfrak A}_{1}[L-t]}\rangle\rangle_{P}$)
in Eq.\ (\ref{eq:integ-pres}) explicitly 
and obtain all-order estimates 
for the expanded quantity under consideration.
Unfortunately, only the first few $d_{n}$ coefficients are known
for most relevant processes.
However, the asymptotic tail 
calculated first by Lipatov et al.~\cite{Lip76},
see for review~\cite{KS80},
allows us to construct 
a model of the generating function $P(t)$  
with reasonable accuracy:
\begin{eqnarray}
 \label{eq:d_n.BLM}
 \tilde{d}_n
  \sim \Gamma(n+1)~n^\gamma c^{n}
       \left[1+O\left(\frac{1}{n}\right)\right]
  \to P(t) \sim
 (t/c)^{\gamma +1}e^{-t/c}\,,
\end{eqnarray}
where $\gamma<1$ and $c$ are numerical coefficients.

Now we illustrate this statement 
by the example 
of perturbative coefficients
$\tilde{d}_n$
for the Adler function
\begin{eqnarray}
 D_\text{S}[L]
  &=& d_0
    + d_1\,\sum_{n=1}^{\infty}\tilde{d}_{n}\,
                       \left(\frac{\alpha_s[L]}{\pi}
                       \right)^{n}
  \label{eq:D_S}
\end{eqnarray}
for scalar correlator, cf.~\cite{BCK05,BMS06},
see the first row in Table~\ref{Tab:1}.
To simulate the $n$-dependence of the coefficients $\tilde{d}_n$,
in accordance with recipe (\ref{eq:d_n.BLM}),
we use the model:
\begin{subequations}
\label{eq:Higgs.Model}
\begin{eqnarray}
 \label{eq:Higgs.d_n.Model}
  \tilde{d}_n^\text{H}
   = c^{n-1}\frac{\Gamma (n+1)+\beta\,\Gamma (n)}{1+\beta}\,,
\end{eqnarray}
which is produced by the generating function
\begin{eqnarray}
 \label{eq:Higgs.P(t).Model}
  P_\text{H}(t)
  &=& \frac{(t/c)+\beta}{c\,(1+\beta)}\,e^{-{t/c}}\,.
\end{eqnarray}
\end{subequations}
In the second row of Table~\ref{Tab:1} we show the results
obtained by fitting two known coefficients 
$\tilde{d}_2$ and $\tilde{d}_3$ 
using the model (\ref{eq:Higgs.Model})\footnote{%
Note that $\tilde{d}_1^\text{H}$ is automatically equal to unity.}.
We see that it gives a very good prediction for 
$\tilde{d}_4^\text{H}=662$,
which appears to be quite close to the value 620
calculated by Chetyrkin \textit{et al.} in~\cite{BCK05}
a year ago.
If we use instead the fitting procedure with taking into account
the fourth coefficient $\tilde{d}_4$,
then the parameters  $c$ and $\beta$ of our model (\ref{eq:Higgs.Model})
change slightly: 
$\left\{c=2.5,~\beta=-0.48\right\}$ 
$\to$ 
$\left\{c=2.4,~\beta=-0.52\right\}$.
This shows the reasonability of our modelling.

\begin{table}[h]
\centerline{
\begin{tabular}{|c|ccccc|}\hline \hline
 Model & $\tilde{d}_1\vphantom{^{|}_{|}}$
               &$\tilde{d}_2$ 
                       &$\tilde{d}_3$ &$\tilde{d}_4$ &$\tilde{d}_5$
\\ \hline \hline
~~~pQCD results with $N_f=5$~~~
       & 1     & 7.42  & 62.3 & 620   & --- $\vphantom{^{|}_{|}}$
\\ \hline
Model (\ref{eq:Higgs.Model})
with $c=2.5,~\beta=-0.48$
       & 1     & 7.42  & 62.3 & 662   & 8615 $\vphantom{^{|}_{|}}$
\\ \hline
Model (\ref{eq:Higgs.Model})
with $c=2.4,~\beta=-0.52$
       & 1     & 7.50  & 61.1 & 625   & 7826 $\vphantom{^{|}_{|}}$
\\ \hline
``NNA'' prediction of~\cite{BKM01}
       & 1     & 3.87  & 21.7 & 122   & 1200 $\vphantom{^{|}_{|}}$
\\ \hline \hline
\end{tabular}}
\caption{Coefficients $\tilde{d}_n$ for Higgs boson decay series.}
\label{Tab:1}
\end{table}

Predictions for $\tilde{d}_n$ values in the so-called 
``Naive Non-Abelianization'' (NNA) approximation~\cite{BKM01}
are presented in the fourth line of Table~\ref{Tab:1}
and they appear to be significantly smaller
than the exact values as well as our model estimates.

\subsection{Series summation in the global APT}
\label{subsec:APT.Global}
Consider the situation with resummation in the global APT
when we take into account heavy-quark thresholds.
We consider here the following series:
\begin{eqnarray}
 {\cal R}^\text{\tiny glob}[L]
  &=& d_0
    + d_1\,\sum_{n=1}^{\infty}\tilde{d}_{n}\,
            {\mathfrak A}^\text{\tiny glob}_{n}[L]\,.
  \label{eq:R.Global}
\end{eqnarray}
Note that due to different normalizations
of ${\mathfrak A}_{n}[L]$
and ${\mathfrak A}^\text{\tiny glob}_{n}[L]$ 
the coefficients $\tilde{d}_{n}$ 
in Eqs.\ (\ref{eq:Ini.Series}) and (\ref{eq:R.Global})
are different.
In Appendix~\ref{App:Threshold.MAPT} we consider in detail
the case with only one threshold,
corresponding to the transition $N_f=3\to N_f=4$,
and provide the result of summation in the Minkowski region 
in Eqs.\ (\ref{eq:sum.R.Glo.4})--(\ref{eq:Delta.f.Mink}).
With taking into account all other thresholds 
the final result reads 
\begin{eqnarray}
 {\cal R}^\text{\tiny glob}[L] 
  =  d_0    
  \!\!&\!+\!&\!\! d_1\sum\limits_{f=3}^{6}
         \theta\left(L_{f}\!\leq\!L\!<\!L_{f+1}\right)
          \langle\langle{\bar{\mathfrak A}_{1}\!\Big[L\!+\!\lambda_f\!-\!\frac{t}{\beta_f};f\Big]
              }\rangle\rangle_{P}\nonumber\\
  \!\!&\!+\!&\!\! d_1\sum\limits_{f=3}^{5} 
         \theta\left(L_{f}\!\leq\!L\!<\!L_{f+1}\right)
          \sum\limits_{k=f+1}^{6}
          \langle\langle{\Delta_{k}\bar{\mathfrak A}_{1}[t]
              }\rangle\rangle_{P}\,,~~~
 \label{eq:sum.R.Glo.456} 
\end{eqnarray}
where $\lambda_f\equiv\ln(\Lambda_3^2/\Lambda_f^2)$ 
describes the shift of the logarithmic argument
due to the change of the QCD scale parameter ($\Lambda_3\to\Lambda_f$),
$L_{f}\equiv\ln(m_f^2/\Lambda_3^2)$,
$L_3=-\infty$ and $L_7=+\infty$.
The second term in Eq.\ (\ref{eq:sum.R.Glo.456}) looks 
like a natural generalization 
of the fixed-$N_f$ formula (\ref{eq:integ-pres})
with taking into account different QCD scales $\Lambda_f$ 
for each fixed-$N_f$ region,
while the last term appears due to continuity of 
${\cal R}^\text{\tiny glob}[L]$ 
at heavy-quark thresholds. 
Using a toy model $P_\text{toy}(t) = \delta(t-2)$
we estimate a relative contribution of the 
$\sum\limits_{k=f+1}^{6}\langle\langle{\Delta_{k}\bar{\mathfrak A}_{1}[t]}\rangle\rangle_{P}$-term: 
it is of the order of 3\% for the $f=4$ threshold,
0.5\% for the $f=5$ threshold,
and 0.1\% for the $f=6$ threshold.

In the Euclidean region the result of summation, 
as shown in~\cite{AB08}, 
is more complicated:
\begin{eqnarray}
 {\cal D}^\text{\tiny glob}[L] 
   = d_0
\!\!&\!+\!&\!\!
     d_1\,\sum\limits_{f=3}^{6}
            \langle\langle{\int_{L_{f}}^{L_{f+1}}\!
                  \frac{\bar{\rho}_{1}\left[L_{\sigma}+\lambda_f;N_f\right]\,dL_\sigma}
                       {1+e^{L-L_{\sigma}-t/\beta_f}}
                }\rangle\rangle_{P}\nonumber\\
\!\!&\!+\!&\!\!
     d_1\,\sum\limits_{f=4}^{6}
            \langle\langle{\Delta_{f}[L,t]}\rangle\rangle_{P}\,,\
 \label{eq:sum.D.Glo.456}
\end{eqnarray}
where corrections $\langle\langle{\Delta_{f}[L,t]}\rangle\rangle_{P}$
to the naive expectation formula 
are defined as 
\begin{eqnarray}
 \Delta_{f}[L,t]
  &\equiv& 
        \int_{0}^{1}\!
         \frac{\bar{\rho}_{1}\left[L_f+\lambda_f-tx/\beta_f;N_f\right]\,t}
              {\beta_f\,\left[1+e^{L-L_f-t\bar{x}/\beta_f}\right]}\,dx\nonumber\\
    &-& \int_{0}^{1}\!
         \frac{\bar{\rho}_{1}\left[L_f+\lambda_{f-1}-tx/\beta_{f-1};N_{f-1}\right]\,t}
              {\beta_{f-1}\,\left[1+e^{L-L_f-t\bar{x}/\beta_{f-1}}\right]}\,dx\,.~~~
 \label{eq:Delta.f.Eucl}
\end{eqnarray}
and now, in contrast to the Minkowski case, 
they explicitly depend on $L$.
To estimate the effect of the thresholds in the Euclidean domain,
we compare the numeric value of 
(\ref{eq:sum.D.Glo.456})
for our toy model $P_\text{toy}(t) = \delta(t-2)$
with the numeric value of the `naive' summation formula
\begin{eqnarray}
 {\cal D}^\text{nai}[L] 
   = d_0  
   + d_1\,\sum\limits_{f=3}^{6}
            \int_{L_{f}}^{L_{f+1}}\!
             \langle\langle{\frac{\bar{\rho}_{1}\left[L_{\sigma}+\lambda_f;N_f\right]\,dL_\sigma}
                  {1+e^{L-L_{\sigma}-t/\beta_f}}
                  }\rangle\rangle_{P_\text{toy}(t)}\,.
 \label{eq:sum.D.nai} 
\end{eqnarray}
The relative values of the differences 
between (\ref{eq:sum.D.Glo.456}) and (\ref{eq:sum.D.nai}) 
are of the order of 2.5\% 
at $L\sim5$ and are about 0.5\% for $L\gtrsim12$.

\section{Fixed-$N_f$ and global FAPT}
\label{sec:FAPT.Summation}

Sometimes perturbative series starts not with the unity term
like $d_0$ in Eq.\ (\ref{eq:Ini.Series})
but rather with $d_0\,a^{\nu}[L]$ 
with power $\nu$ being a fractional number,
see, for example, \cite{BMS06}.
Then one needs to use FAPT 
in order 
to obtain an analytic image of the series.
For this reason 
we need to generalize the APT summation method
discussed in section~\ref{sec:APT.Summation}
in order to apply it in FAPT case.

\subsection{Resummation of expansions in the fixed-$N_f$ FAPT}
\label{subsec:FAPT.fixed-Nf}
First, we consider the case of fixed $N_f$.
We start with a power series in 
the standard perturbative QCD
\begin{eqnarray}
 D_{\nu}[L]
  &=& d_0\,a^{\nu}[L]
    + d_1\,\sum_{n=1}^{\infty}\tilde{d}_{n}\,
                       a^{n+\nu}[L]\,,
  \label{eq:R-PT_nu0}
\end{eqnarray}
with $\nu>0$ that appears due to renormgroup 
summation.
This type of series
generates the following analytic image 
in the Minkowski region:
\begin{eqnarray}
 {\cal R}_{\nu}[L]
  &=& d_0\,{\mathfrak A}_{\nu}[L]
    + d_1\,\sum_{n=0}^{\infty}\tilde{d}_{n+1}\,
            {\mathfrak A}_{n+1+\nu}[L]\,.
  \label{eq:R-MFAPT}
\end{eqnarray}
Due to the recurrence relation
\begin{eqnarray}
 \label{eq:Rec.Rel.n+1+nu}
 {\mathfrak A}_{n+\nu}[L]
  = \frac{\Gamma(\nu)}{\Gamma(n+\nu)}
    \left(-\frac{d}{dL}\right)^{n}{\mathfrak A}_{\nu}[L]
 \end{eqnarray}
we have
\begin{eqnarray}
 \!\!\!{\cal R}_{\nu}[L]
  &=& d_0\,{\mathfrak A}_{\nu}[L]
    + d_1\,\langle\langle{X(t;1+\nu)}\rangle\rangle_{P}\,,
 \label{eq:Q1}
\end{eqnarray}
where
\begin{eqnarray}\label{eq:X(a)}
 \label{eq:A1}
  X(t;1+\nu) &\equiv &
  \sum_{n=0}^{\infty}
   \frac{(-\hat{x}_t)^n\Gamma(1+\nu)}{\Gamma(n+1+\nu)}\,
    {\mathfrak A}_{1+\nu}[L]
  \text{~~~with~~~} \hat{x}_t=t\,\frac{d}{dL}\,.
\end{eqnarray}
We have a nice representation
for the ratio of $\Gamma$-functions
(see formula 5.2.7.20 in~\cite{PBM-EF81})
\begin{eqnarray}
 \label{eq:Gamma.ratio}
  \frac{\Gamma(n+1)\Gamma(\nu+1)}{\Gamma(n+\nu+1)}
  & =& \int_0^1\!\left(1-u^{1/\nu}\right)^{n}du\,,
\end{eqnarray}
which gives us a possibility to process further 
with our series (\ref{eq:A1})
\begin{eqnarray*}
 \label{eq:Sum.Gamma.ratio}
  \sum_{n=0}^{\infty}
   \frac{(-\hat{x}_t)^n\Gamma(1+\nu)}{\Gamma(n+1+\nu)}
   = \int_0^{1}\!
      \exp\left(\hat{x}_t\cdot u^{1/\nu}-\hat{x}_t\right)du\,.
\end{eqnarray*}
Now we recall that $\hat{x}_t$ is the operator $\hat{x}_t=t\cdot{d}/{dL}$.
That means that the operator  $e^{z\,\hat{x}_t}$
when acting upon a function $A[L]$ just shifts its argument:
$e^{z\,\hat{x}_t}A[L]=A[L+tz]$,
and we have
\begin{eqnarray}
 X(t;1+\nu)
  = \int_0^{1}\!
     {\mathfrak A}_{1+\nu}
      \left[L+t\left(u^{1/\nu}-1\right)\right]
       du\,.
 \label{eq:X.A}
\end{eqnarray}
Substituting Eq.\ (\ref{eq:X.A}) in Eq.\ (\ref{eq:Q1}),
we obtain as a result:
\begin{eqnarray}
 {\cal R}_{\nu}[L]
  &=&d_0\,{\mathfrak A}_{\nu}[L]
   + d_1\,\int_0^{1}\!\!du
      \langle\langle{{\mathfrak A}_{1+\nu}
           \left[L-t\left(1-u^{1/\nu}\right)\right]
           }\rangle\rangle_{P}\nonumber
 \\
  &=&d_0\,{\mathfrak A}_{\nu}[L]
   + d_1\,\langle\langle{{\mathfrak A}_{1+\nu}[L-t]
         }\rangle\rangle_{P_{\nu}}\,,
 \label{eq:final.nu.to.0}
\end{eqnarray}
where
\begin{subequations}
\begin{eqnarray}
 P_{\nu}(t)\!\!
  &\!\equiv\!&\!\!
   \int_0^{1}\!\!P\left(\frac{t}{1-u^{1/\nu}}\right)
    \frac{du}{1-u^{1/\nu}}
   = \int_0^{1}\!\!P\left(\frac{t}{1-x}\right)
        \Phi_{\nu}(x)\frac{dx}{1-x}\,;~~~
 \label{eq:P.nu}\\
 \Phi_{\nu}(x)\!\!
  &\!\equiv\!&\!\! \nu\,x^{\nu-1}
   \mathop{\longrightarrow}\limits_{\nu\to0+}
    \delta(x)\,.
 \label{eq:Phi.nu.0}
\end{eqnarray}
\end{subequations}
Note here that $\lim\limits_{\nu\to0+}P_{\nu}(t)=P(t)$.

A similar formula can be obtained to sum up 
perturbative series in the Euclidean domain:
\begin{eqnarray}
 {\cal D}_{\nu}[L]
  &=&d_0\,{\cal A}_{\nu}[L]
   + d_1 \langle\langle{{\cal A}_{1+\nu}[L-t]}\rangle\rangle_{P_{\nu}}\,,
 \label{eq:Sum.Integ.Repr.FAPT.Eucl}         
\end{eqnarray} 

We see that the main difference between FAPT and APT 
with respect to a series summation technique
is that in FAPT one needs 
to modify the initial generating function $P(t)$
into a new one, $P_{\nu}(t)$.
It is interesting to note that for our model 
\begin{eqnarray}
 \label{eq:d.P.factorial}
  \tilde{d}_n &=& c^{n-1}\frac{\Gamma(n+\delta)}{\Gamma(1+\delta)}\,,\quad
  P(t;\delta)\ =\ \left(\frac{t}{c}\right)^\delta e^{-t/c}
\end{eqnarray}
the integration for $P_{\nu}(t)$ in Eq.\ (\ref{eq:P.nu})
can be carried out exactly to produce rather a complicated expression
containing the regularized confluent hypergeometric function.
For integer values of $\delta=m\geq0$ formula is simplified
to 
\begin{eqnarray}
 \label{eq:P.nu.factorial.m}
  P_{\nu}(t;m)
  = \Gamma (\nu+1)\,
     G_{1,2}^{2,0}
      \left(\frac{t}{c}\left|
       \begin{array}{c}\nu\\0,m\end{array}
             \right.
      \right),
\end{eqnarray}
where $G_{1,2}^{2,0}(z|...)$ is the Meijer $G$-function
defined as
\begin{eqnarray}
 G_{1,2}^{2,0}
      \left(z
            \left|
             \begin{array}{c} a \\ 
                             {b_1,b_2}
             \end{array}
            \right.
      \right)
 = \frac{1}{2\pi\,i}
    \oint_{C}
     \frac{\Gamma(b_1+s)\Gamma(b_2+s)}{\Gamma(a+s)}\,
      \frac{ds}{z^s}\,,
\end{eqnarray}
where the contour $C$ of integration 
is set up so 
that the poles of $\Gamma(b_1+s)$ and $\Gamma(b_2+s)$
lie in the same region (internal or external) 
with respect to the contour $C$. 

\subsection{Resummation of expansions in the global FAPT}
\label{subsec:FAPT.global}

To generalize formulae obtained in the previous 
subsection to the case of the global FAPT,
we follow the lines of subsection~\ref{subsec:APT.Global}
and take into account the new recurrence relation (\ref{eq:Rec.Rel.n+1+nu})
for couplings in FAPT.
We start with the analytic image of perturbative series 
in the Minkowski region
\begin{eqnarray}
 {\cal R}_{\nu}^\text{\tiny glob}[L]
  &=& d_0\,{\mathfrak A}_{\nu}^\text{\tiny glob}[L]
    + d_1\,\sum_{n=0}^{\infty}\tilde{d}_{n+1}\,
            {\mathfrak A}_{n+1+\nu}^\text{\tiny glob}[L]
 \label{eq:R-MFAPT.Global}
\end{eqnarray}
and proceeding carefully 
obtain the result \footnote{%
With the same notation $L_3=-\infty$ and $L_7=+\infty$
as in Eqs.\ (\ref{eq:sum.R.Glo.456}) and (\ref{eq:sum.D.Glo.456}).}:
\begin{eqnarray}
 {\cal R}_{\nu}^\text{\tiny glob}[L]
   = d_0\,{\mathfrak A}^\text{\tiny glob}_{\nu}[L]
\!\!&\!\!+\!\!&\!\!
     d_1\sum\limits_{f=3}^{6}
         \theta\left(L_{f}\!\leq\!L\!<\!L_{f+1}\right)
          \langle\langle{\bar{\mathfrak A}_{1+\nu}\!\Big[L\!+\!\lambda_f\!-\!\frac{t}{\beta_f};f\Big]
                        }\rangle\rangle_{P_{\nu}}
 \nonumber\\
\!\!&\!\!+\!\!&\!\!
    d_1\sum\limits_{f=3}^{6}
         \theta\left(L_{f}\!\leq\!L\!<\!L_{f+1}\right)\!\!
           \sum\limits_{k=f+1}^{6}\!
            \langle\langle{\Delta_{k}\bar{\mathfrak A}_{1+\nu}[t]}\rangle\rangle_{P_{\nu}}.~
 \label{eq:Glo-MFAPT.sum.R.Glo.456} 
\end{eqnarray}
Formulae for the Euclidean region appear to be more complicated
and we refer the interested reader to the review~\cite{AB08},
where this case is analyzed and all formulas
are derived.

\section{Application to the Higgs boson decay}
\label{sec:Appl.Higgs}
In order to apply the developed techniques
to the estimation of real processes,
we need first to model the generating function $P(t)$ 
of the corresponding perturbative coefficients.
As we know from the discussion in section~\ref{sec:APT.Summation},
in the case of the Higgs boson decay
$H^0\to\bar{b}b$
the model (\ref{eq:Higgs.Model})
is of reasonable accuracy.

First, we recall 
what is the object of interest 
in the case of the Higgs boson decay to a $\bar{b}b$ pair.
Here we have for the decay width 
\begin{eqnarray}
 \Gamma(\text{H} \to b\bar{b})
  = \frac{G_F}{4\sqrt{2}\pi}\,
     M_{H}\,
      \widetilde{R}_\text{S}(M_{H}^2)\,,
 \label{eq:Higgs.decay.rate}
\end{eqnarray}
where 
$\widetilde{R}_\text{S}(M_{H}^2)$
is just $m^2_{b}(M_{H}^2)\,R_\text{S}(M_{H}^2)$.
In the one-loop FAPT this generates the following
non-power expansion\footnote{%
Appearance of denominators $\pi^n$ in association 
with the coefficients $\tilde{d}_n$
is a consequence of $d_n$ normalization,
see Eq.\ (\ref{eq:D_S}).}: 
\begin{eqnarray}
 \widetilde{R}_\text{S}^{\text{FAPT}}[L]
   =  3\,\hat{m}_{(1)}^2\,
      \left[{\mathfrak A}_{\nu_{0}}^{\text{\tiny glob}}[L]
          + d_1\,\sum_{n\geq1} 
             \frac{\tilde{d}_{n}}{\pi^{n}}\,
              {\mathfrak A}_{n+\nu_{0}}^{\text{\tiny glob}}[L]    
      \right]\,,
 \label{eq:R_S-MFAPT} 
\end{eqnarray}
where $\hat{m}_{(1)}^2$ is the renormalization-group
invariant of the one-loop $m^2_{b}(\mu^2)$ evolution
\begin{eqnarray}
 \label{eq:m2-hat-run}
  m_{b}^2(Q^2)
   = \hat{m}_{(1)}^2\,
      \alpha_{s}^{\nu_{0}}(Q^2)
\end{eqnarray} 
with $\nu_{0}=2\gamma_0/b_0(5)=1.04$ and 
$\gamma_0$ is the quark-mass anomalous dimension.

We take the model (\ref{eq:Higgs.Model}) 
and apply Eq.\ (\ref{eq:Glo-MFAPT.sum.R.Glo.456})
to estimate
how good is FAPT 
in approximating the whole sum $\widetilde{R}_\text{S}^{\text{FAPT}}[L]$
in the range $L\in[11,13.8]$
which corresponds to the range
$M_H\in[60,170]$~GeV$^2$ 
with $\Lambda^{N_f=3}_{\text{QCD}}=172$~MeV
and ${\mathfrak A}^{\text{\tiny glob}}_{1}(m_Z^2)=0.120$.
In this range we have $L_5<L<L_6$,
so that Eq.\ (\ref{eq:Glo-MFAPT.sum.R.Glo.456})
transforms into
\begin{eqnarray}
 \frac{\widetilde{R}_\text{S}^{\text{FAPT}}[L]}
      {3\,\hat{m}_{(1)}^2}
   = {\mathfrak A}^\text{\tiny glob}_{\nu_{0}}[L]
   +  \frac{d_1}{\pi}\,
       \langle\langle{\bar{\mathfrak A}_{1+\nu_{0}}\!\Big[L\!+\!\lambda_5\!-\!\frac{t}{\pi\beta_5};5\Big]
                     \!+\!\Delta_{6}\bar{\mathfrak A}_{1+\nu_{0}}\left[\frac{t}{\pi}\right]
                      }\rangle\rangle_{P_{\nu_{0}}}
 \label{eq:R_S.Sum} 
\end{eqnarray}
with $P_{\nu_{0}}(t)$ defined via Eqs.\ (\ref{eq:Higgs.Model}) 
and (\ref{eq:P.nu})
with the parameters $c=2.4$, $\beta=-0.52$, and $\nu_{0}=1.04$.

\begin{figure}[t]
 \centerline{\includegraphics[width=0.5\textwidth]{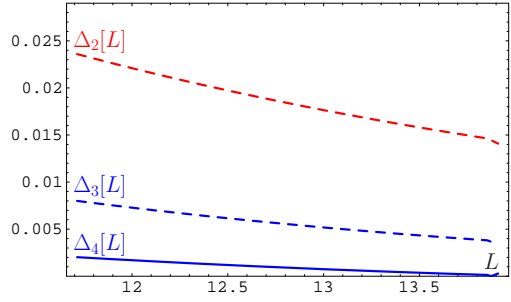}
  \vspace*{-1mm}}
   \caption{The relative errors  $\Delta_N[L]$, $N=2, 3$ and $4$,  
   of the truncated FAPT, Eq.\ (\ref{eq:FAPT.trunc}), as compared to
   the exact summation result, Eq.\ (\ref{eq:R_S.Sum}).
   \label{fig:Higgs}}
\end{figure}

Now we analyze the accuracy of the truncated FAPT expressions
\begin{eqnarray}
 \label{eq:FAPT.trunc}
 \widetilde{R}_\text{S}^{\text{FAPT}}[L;N]
  &=& 3\,\hat{m}_{(1)}^2\,
       \left[{\mathfrak A}_{\nu_{0}}^{\text{\tiny glob}}[L]
           + d_1\,\sum_{n=1}^{N}
              \frac{\tilde{d}_{n}}{\pi^{n}}\,
               {\mathfrak A}_{n+\nu_{0}}^{\text{\tiny glob}}[L]
       \right]
\end{eqnarray}
and compare them with the total sum 
$\widetilde{R}_\text{S}^{\text{FAPT}}[L]$
in Eq.\ (\ref{eq:R_S.Sum})
using relative errors
\begin{eqnarray}
 \label{eq:relat.errors}
  \Delta_N[L]
   &=& 1 
    - \frac{\widetilde{R}_\text{S}^{\text{FAPT}}[L;N]}
           {\widetilde{R}_\text{S}^{\text{FAPT}}[L]}\,.
\end{eqnarray}
In Fig.\ \ref{fig:Higgs}, 
we show these errors for $N=2$, $N=3$, and $N=4$
in the analyzed range of $L\in[11,13.8]$.
We see that already $\widetilde{R}_\text{S}^{\text{FAPT}}[L;2]$
gives accuracy of the order of 2.5\%,
whereas $\widetilde{R}_\text{S}^{\text{FAPT}}[L;3]$
of the order of 1\%.
That means that there is no need to calculate further corrections:
at the level of accuracy of 1\% it is quite enough to take into account 
only coefficients up to $d_3$.
This conclusion is stable
with respect to the variation of parameters
of the model $P_\text{H}(t)$.
These estimates 
demonstrate also a good convergence of the considered series in FAPT.

It is also interesting to estimate how large is the effective shift
$\Delta{L}$
which models the exact result (\ref{eq:R_S.Sum})
by the simple formula
\begin{eqnarray}
 \label{eq:Higgs.Effective}
  \widetilde{R}_\text{S}^{\text{FAPT}}[L]
   \approx 3\,\hat{m}_{(1)}^2\,
    \left[{\mathfrak A}_{\nu_{0}}^{\text{\tiny glob}}[L]
        + \frac{d_1}{\pi(1+\nu_{0})}\,
           {\mathfrak A}_{1+\nu_{0}}^{\text{\tiny glob}}
             \Big[L-\Delta{L}\Big]
   \right]\,.
\end{eqnarray}
Numerical estimation shows that $\Delta{L}\approx2.16$.
We can transform this result 
into a finite shift of $\Lambda^{N_f=3}_\text{QCD}=0.172$~MeV:
\begin{eqnarray}
 \label{eq:Lambda.Shift}
  \Lambda^{N_f=3}_\text{QCD}
  \to 
  \Lambda^{N_f=3}_\text{eff} 
   = 2.94\,
     \Lambda^{N_f=3}_\text{QCD}
   \simeq 505~\text{MeV}\,.
\end{eqnarray}
If one estimate this shift by using only the $d_2$-term,
then the result is 10\% smaller:
$\Lambda^{N_f=3}_\text{eff}\simeq 450$~MeV.

\section{Conclusions}
In this paper,
using recurrence relations of the one-loop (F)APT,
see Eqs.\ (\ref{eq:recurrence}) and (\ref{eq:Rec.Rel.n+1+nu}),
we construct all needed rules 
to sum up nonpower series expansions
in the fixed-$N_f$ and global (F)APT.
We show how these results can be used 
to estimate the rate of convergency of nonpower series expansion.
In the case of the series
appearing in the QCD description of the Higgs boson decay 
into a $\bar{b}b$-pair, 
this convergence is estimated to be moderately fast:
in order to have 1\%-level of accuracy, 
it is sufficient to sum up the first four terms 
($d_0$, $d_1$, $d_2$, and $d_3$).
In this case, such summation is equivalent 
in the considered range of $L\in[11,13.8]$
to the constant shift of $\Lambda^{N_f=3}_\text{QCD}$
from 172~MeV to 505~MeV.

\section*{Acknowledgments}
We would like to thank A.~L.~Kataev, A.~V.~Sidorov, 
D.~V.~Shirkov, and O.~P.~Solovtsova,  
for stimulating discussions, useful remarks, 
and clarifying criticism.
We are thankful to N.~G.~Stefanis for the long-run fruitful collaboration
on the (F)APT, 
moreover S.V.M. thanks N.G.S. 
for the discussing and clarifying the results of~\cite{MS04}.
We are indebted to Prof.\ Klaus Goeke and Dr. Dr. Nicos Stefanis
for the warm hospitality 
at Bochum University 
where this work was partially prepared.
This work was supported in part by the Heisenberg--Landau Programme,
grants 2005--2008, 
and the Deutsche Forschungsgemeinschaft
(Project DFG 436 RUS 113/881/0-1),
that allows us to collaborate with Bochum group,
and by the Russian Foundation for Fundamental Research, grants
No.\ 
06-02-16215, 07-02-91557, and 08-01-00686,
and the BRFBR--JINR Cooperation Programme, contract No.\ F06D-002.

\begin{appendix}
\appendix
\makeatletter
\@addtoreset{equation}{section}%
 \def\theequation@prefix{\thesection}%
 \def\theequation{\theequation@prefix\arabic{equation}}%
\makeatother
\section{Single quark threshold in the Minkowski APT}
 \label{App:Threshold.MAPT}
We consider here only one heavy-quark threshold
corresponding to the transition $N_f=3\to N_f=4$.
In this case, the global one-loop spectral density 
$\rho^\text{\tiny glob}_n(s)=\rho^\text{\tiny glob}_n[L(s)]$
(with $L(s)=\ln\left(s/\Lambda_3^2\right)$)
is expressed in terms of fixed-flavor spectral densities 
with 3 and 4 flavors,
$\bar{\rho}_n[L;3]$ and $\bar{\rho}_n[L+\lambda_4;4]$,
see Eq.\ (\ref{eq:rho_n.Glo}): 
\begin{eqnarray}       
 \label{eq:global_PT_Rho_4}
  \rho_n^\text{\tiny glob}[L]  
  &=& \theta\left(L<L_{4}\right)\,
       \bar{\rho}_n\left[L;3\right]
    + \theta\left(L_{4}\leq L\right)\,
       \bar{\rho}_n\left[L+\lambda_4;4\right]\,,~~~
\end{eqnarray}
with $\lambda_4\equiv\ln\left(\Lambda_3^2/\Lambda_4^2\right)$ 
describing a shift of the logarithmic argument 
due to the change of the QCD scale parameter ($\Lambda_3\to\Lambda_4$),
$L_{4}\equiv\ln\left(m_4^2/\Lambda_3^2\right)$,
and $\bar{\rho}_{n}\left[L;N_f\right]$ 
defined in Eq.\ (\ref{eq:Glo.Couplings}).
In the one-loop approximation we have a very useful property
of the spectral densities $\rho_{n}\left[L\right]$
and $\bar{\rho}_{n}\left[L;N_f\right]$:
\begin{subequations}
\label{eq:Reccur.SD.n+1.1}
\begin{eqnarray}
 \label{eq:Reccur.Nf.SD.n+1.n}
  \rho_{n+1}[L]
   &=& \frac{1}{n}
        \left(-\frac{d}{dL}\right)
         \rho_{n}[L]\
    =\ \frac{1}{\Gamma(n+1)}
        \left(-\frac{d}{dL}\right)^n
         \rho_{1}[L]\,;
\\
 \label{eq:Reccur.Nf.bar.SD.n+1.n}
  \bar{\rho}_{n+1}[L;N_f]
   &=& \frac{1}{n\,\beta_f}
        \left(-\frac{d}{dL}\right)
         \bar{\rho}_{n}[L;N_f]\\
   &=& \frac{1}
            {\Gamma(n+1)\,\beta_f^{n+1}}
        \left(-\frac{d}{dL}\right)^n
         \rho_1[L]\,,
 \label{eq:Reccur.Nf.bar.SD.n+1.1}
\end{eqnarray}
\end{subequations}
which is valid for $n\geq0$ (except only Eq.\ (\ref{eq:Reccur.Nf.bar.SD.n+1.n})) 
and allows us immediately to rewrite Eq.\ (\ref{eq:U.n.Glo})
in a more explicit form,
where all $n$-dependencies appear explicitly:
\begin{eqnarray}
  {\mathfrak A}_{n+1}^\text{\tiny glob}[L]
  &=& \frac{\theta\left(L<L_{4}\right)}
           {\Gamma(n+1)}
       \Big\{\frac{1}{\beta_3^n}
              \left[\left(-\frac{d}{dL}\right)^{n}
                     \bar{\mathfrak A}_{1}[L;3]
                  - \left(-\frac{d}{dL_{4}}\right)^{n}
                     \bar{\mathfrak A}_{1}[L_{4};3]
              \right] \nonumber\\ 
  & &~~~~~~~~~~~~~+ 
             \frac{1}{\beta_4^n}\,
              \left(-\frac{d}{dL_{4}}\right)^{n}
               \bar{\mathfrak A}_{1}[L_{4}+\lambda_{4};4]
       \Big\}\nonumber\\ 
 \label{eq:Reccur.Glo.U.n+1.1}
  &+& \frac{\theta\left(L\geq L_{4}\right)}
           {\Gamma(n+1)}
       \frac{1}{\beta_4^n}
        \left(-\frac{d}{dL}\right)^n
         \bar{\mathfrak A}_{1}[L+\lambda_{4};4]\,.
\end{eqnarray}
Thus the general structure of the $n$-dependence,
which we have in Eq.\ (\ref{eq:Reccur.Glo.U.n+1.1}),
is $x^n/\Gamma(n+1)$ with $x=[1/\beta_f](-d/dL)$
and we know how to sum up this structure, see Eq.\ (\ref{eq:integ-pres}):
\begin{eqnarray}
 {\cal R}^\text{\tiny glob}[L]\!\!\!
   &\!\equiv\!&\!\!\! d_0 
          + \sum_{n=0}^{\infty}
             d_{n+1}\,{\mathfrak A}^\text{\tiny glob}_{n+1}[L]
    \equiv  d_0 
          + d_1\,\sum_{i}\,\theta_i[L]\,{\mathfrak S}_{f;i}[L+\lambda_f]\,;~~~~~
 \\
 {\mathfrak S}_{f;i}[L]\!\!\!
   &\!\sim\!&\!\!\!
          \sum_{n=0}^{\infty}
            \frac{\langle\langle{t^n}\rangle\rangle_{P}}{\beta_f^n\,\Gamma(n+1)}
             \left(\frac{-d}{dL}\right)^n
              \bar{\mathfrak A}_{1}[L;N_f]
   \!=\! \langle\langle{\bar{\mathfrak A}_{1}\!\left[L-\frac{t}{\beta_f};N_f\right]}
       \rangle\rangle_{P}\,.~~~~~
 \label{eq:sum.Sigma}
\end{eqnarray}
Collecting all types of different $\theta$-structures in Eq.\ (\ref{eq:Reccur.Glo.U.n+1.1})
and inserting them in Eq.\ (\ref{eq:sum.Sigma})
we arrive at the answer
\begin{eqnarray}
 ~~~{\cal R}^\text{\tiny glob}[L] 
  = d_0\!\!\!  
  &\!+\!&\!\!\! d_1\,\theta\left(L<L_{4}\right)
        \langle\langle{\bar{\mathfrak A}_{1}\!\Big[L-\frac{t}{\beta_3};3\Big]
           + \Delta_{4}\bar{\mathfrak A}_{1}[t]
            }\rangle\rangle_{P}   
  \nonumber\\
 \!\!\!
  &\!+\!&\!\!\! d_1\,\theta\left(L\geq L_{4}\right)
        \langle\langle{\bar{\mathfrak A}_{1}\!\Big[L+\lambda_4-\frac{t}{\beta_4};4\Big]
            }\rangle\rangle_{P}\,,~~~~~~~    
 \label{eq:sum.R.Glo.4}
\end{eqnarray}
where we denoted (with $\lambda_3=0$)
\begin{eqnarray}
 \Delta_{f}\bar{\mathfrak A}_{n}[t] 
  \equiv
     \bar{\mathfrak A}_{n}\!\Big[L_{f}+\lambda_{f}-\frac{t}{\beta_f};N_f\Big]
   - \bar{\mathfrak A}_{n}\!\Big[L_{f}+\lambda_{f-1}-\frac{t}{\beta_{f-1}};N_{f-1}\Big]\,.
 \label{eq:Delta.f.Mink}
\end{eqnarray}

\section{Numerical details}
\label{App:Numbers}
We use the following heavy-quark masses:
$m_{c}=1.2$ GeV, $m_{b}=4.3$ GeV, and
$m_{t}=175$ GeV
that generate the following set of QCD scales 
and logarithms 
$\lambda_f=\ln\left[\Lambda_3^2/\Lambda_f^2\right]$
in the one-loop approximation:
\begin{eqnarray}
 \label{eq:Lambdas.3-6}
  \Lambda_3\!&\!=\!&\!172~\text{MeV},~ 
  \Lambda_4 = 147~\text{MeV},~
  \Lambda_5 = 111~\text{MeV},~
  \Lambda_6 =  55~\text{MeV}\,;~~~ 
 \\
  \lambda_4\!&\!=\!&\!0.311\,,~~~
  \lambda_5 = 0.898\,,~~~
  \lambda_6 = 2.30\,,~~~
  \lambda_Z = \ln\frac{m_Z^2}{\Lambda_3^2}
            = 12.55\,.       
 \label{eq:lambdas.4-6}
\end{eqnarray}
\end{appendix}


\begin{thebibliography}{10}

\bibitem{JS95-349}
 H.~F. Jones and I.~L. Solovtsov,
  Phys. Lett. \textbf{B349},  519  (1995).

\bibitem{JS95-357}
 H.~F. Jones, I.~L. Solovtsov, and O.~P. Solovtsova,
  Phys. Lett. \textbf{B357},  441  (1995).

\bibitem{SS96}
 D.~V. Shirkov and I.~L. Solovtsov,
  JINR Rapid Commun. \textbf{2[76]},  5  (1996)
;\\
  Phys. Rev. Lett. \textbf{79},  1209  (1997).

\bibitem{MS96}
 K.~A. Milton and I.~L. Solovtsov,
  Phys. Rev. \textbf{D55},  5295  (1997).

\bibitem{MiSol97}
 K.~A. Milton and O.~P. Solovtsova,
  Phys. Rev. \textbf{D57},  5402  (1998).

\bibitem{SS98}
 I.~L. Solovtsov and D.~V. Shirkov,
  Phys. Lett. \textbf{B442},  344  (1998).

\bibitem{Mag99}
 B.~A. Magradze,
  Int. J. Mod. Phys. \textbf{A15},  2715  (2000);
  \uppercase{JINR} preprint E2-2000-222, 2000 [hep-ph/0010070];
 \uppercase{P}reprint RMI-2003-55, 2003 [hep-ph/0305020].

\bibitem{KM01}
 D.~S. Kourashev and B.~A. Magradze,
  \uppercase{P}reprint RMI-2001-18, 2001 [hep-ph/0104142];
  Theor. Math. Phys. \textbf{135},  531  (2003).
   
\bibitem{Mag05}
 B.~A. Magradze,
  Few Body Syst. \textbf{40},  71  (2006).

\bibitem{SSK99}
 N.~G. Stefanis, W. Schroers, and H.-C. Kim,
  Phys. Lett. \textbf{B449},  299  (1999);
  Eur. Phys. J. \textbf{C18},  137  (2000).

\bibitem{BPSS04}
 A.~P. Bakulev, K. Passek-Kumeri\v{c}ki, W. Schroers, and N.~G. Stefanis,
  Phys. Rev. \textbf{D70},  033014  (2004); 079906(E)  (2004).

\bibitem{SS06}
 D.~V. Shirkov and I.~L. Solovtsov,
  Theor. Math. Phys. \textbf{150},  132  (2007).

\bibitem{KS01}
 A.~I. Karanikas and N.~G. Stefanis,
  Phys. Lett. \textbf{B504},  225  (2001);
  Erratum-ibid.: \textbf{B636}, 330  (2006).

\bibitem{Ste02}
 N.~G. Stefanis,
  Lect. Notes Phys. \textbf{616},  153  (2003).

\bibitem{BMS05}
 A.~P. Bakulev, S.~V. Mikhailov, and N.~G. Stefanis,
  Phys. Rev. \textbf{D72},  074014  (2005); 119908(E)  (2005).

\bibitem{BKS05}
 A.~P. Bakulev, A.~I. Karanikas, and N.~G. Stefanis,
  Phys. Rev. \textbf{D72},  074015  (2005).

\bibitem{BMS06}
 A.~P. Bakulev, S.~V. Mikhailov, and N.~G. Stefanis,
  Phys. Rev. \textbf{D75},  056005  (2007).

\bibitem{MS04}
 S.~V. Mikhailov,
  JHEP \textbf{06}, 009 (2007).

\bibitem{SM94}
 D.~V. Shirkov and S.~V. Mikhailov,
  Z. Phys. \textbf{C63},  463  (1994).

\bibitem{Rad82}
 A.~V. Radyushkin,
  JINR Rapid Commun. \textbf{78},  96  (1996).

\bibitem{Shi98}
 D.~V. Shirkov,
  Theor. Math. Phys. \textbf{119},  438  (1999).

\bibitem{BRS00}
 A.~P. Bakulev, A.~V. Radyushkin, and N.~G. Stefanis,
  Phys. Rev. \textbf{D62},  113001  (2000).

\bibitem{Lip76}
 L.~N. Lipatov,
  Sov. Phys. JETP \textbf{45},  216  (1977).

\bibitem{KS80}
 D.~I. Kazakov and D.~V. Shirkov,
  Fortsch. Phys. \textbf{28},  465  (1980).

\bibitem{BCK05}
 P.~A. Baikov, K.~G. Chetyrkin, and J.~H. K{\"u}hn,
  Phys. Rev. Lett. \textbf{96},  012003  (2006).

\bibitem{BKM01}
 D.~J.~Broadhurst, A.~L.~Kataev, and C.~J.~Maxwell,
  Nucl. Phys. \textbf{B592}, 247 (2001).

\bibitem{AB08}
 A.~P. Bakulev,
  ``\textit{Fractional Analytic Perturbation Theory in QCD
       with Selected Applications}'',
  to be published in 
  \textsl{Physics of Particles and Nuclei}.
  

\bibitem{PBM-EF81}
 A.~P. Prudnikov, Y.~A. Brychkov, and O.~I. Marichev,
  {\em Integrals and seria. Elementary functions},
  edited by T.~N. Kuznetsova and E.~V. Shikin
  (Nauka, Moscow, 1981). 

\end{thebibliography}

\end{document}